\documentclass[%
 amsmath,amssymb,
 aps,
]{revtex4-1}
\usepackage{color}
\usepackage[utf8]{inputenc}
\usepackage{graphicx}
\usepackage{dcolumn}
\usepackage{bm}

\begin{document}
\title{Mean-field approach for frequency synchronization in complex networks of two oscillator types}
\author{Stefan Wieland}
\email{stefan.wieland@tuta.io}
\author{Simone Blanco Malerba}%
\author{S\'ebastien Aumaitre}%
\author{Herv\'e Bercegol}
\affiliation{SPEC, CEA, CNRS, Universit\'e Paris-Saclay, F-91191 Gif-sur-Yvette Cedex, France}
\begin{abstract}
Oscillator networks with an asymmetric bipolar distribution of natural frequencies are useful representations of power grids. We propose a mean-field model that captures the onset, form and linear stability of frequency synchronization in such oscillator networks. The model takes into account a broad class of heterogeneous connection structures and identifies a functional form as well as basic properties that synchronized regimes possess classwide. The framework also captures synchronized regimes with large phase differences that commonly appear just above the critical threshold. Additionally, the accuracy of mean-field assumptions can be gauged internally through two model quantities. With our framework, the impact of local grid structure on frequency synchronization can be systematically explored.
%
\end{abstract}
\maketitle
\section{Introduction}\label{s:intro}
Synchronization is a frequent phenomenon in nature, appearing in biological, ecological, sociological, and engineering contexts (see \cite{acebron_kuramoto_2005,arenas_synchronization_2008,dorfler_synchronization_2014,gupta_kuramoto_2014,rodrigues_kuramoto_2016} for surveys). These systems can be considered networks of coupled phase oscillators and are described by the paradigmatic Kuramoto model \cite{kuramoto_chemical_1984} or its extensions. In this setting, the phase $\theta_j$ of oscillator $j$ is driven both by its \emph{natural} frequency $\omega_j$ and by the phase difference with connected oscillators that couple with uniform strength $\lambda$. The connection structure is encoded in the adjacency matrix with entries $A_{jn}=1$ if oscillators $j$ and $n$ are connected and $A_{jn}=0$ otherwise. The \emph{actual} frequency of oscillator $j$ in a  network of $N$ oscillators is then given by
\begin{equation}\label{e:kuramoto} 
\dot{\theta}_j=\omega_j+\lambda\sum_{n=1}^{N} A_{jn} \sin{(\theta_n-\theta_j)} \, .
\end{equation}
Due to the symmetric coupling, the system frequency $\sum_{j=1}^{N}\dot{\theta}_j/N=\sum_{j=1}^{N}\omega_j/N$ is constant. In literature, natural frequencies are commonly shifted to fulfill $\sum_{j=1}^{N}\omega_j=0$ without loss of generality, entering a reference frame that co-rotates with the system frequency. Moreover, one usually assumes that the network is connected, i.e., that along network links, each oscillator can be reached from any other. 

If all natural frequencies are finite and the coupling strength surpasses a critical threshold $\lambda^*$, all oscillator frequencies eventually attain the same value - this ordered regime is called frequency synchronization \cite{dorfler_synchronization_2014,kuramoto_statistical_1987,strogatz_kuramoto_2000}. In the co-rotating reference frame, it is described by stable steady states of Eq.~\ref{e:kuramoto} where all oscillator phases are ``locked'', but in general of different value. Each steady state belongs to a continuum of fixed points that differ only by an arbitrary uniform rotation of all oscillator phases. The onset and form of frequency synchronization is characterized by the critical threshold and the values of locked phases, respectively, both of which depend on $A_{jn}$ (the network topology) and the set $\{\omega_j\}$ of natural frequencies. There are other synchronization regimes that have been extensively explored, most notably a partial synchronization stage at weaker coupling strengths where oscillator frequencies are not constant and equal, but fluctuating and only positively correlated (see for instance \cite{kuramoto_statistical_1987,strogatz_kuramoto_2000,restrepo_onset_2005}). Yet frequency synchronization is the regime of interest in many realistic settings where sparsely coupled oscillators have non-identical natural frequencies \cite{dorfler_synchronization_2014}.

As an example, frequency synchronization is paramount to stable operations of power grids, i.e., networks that transmit and distribute electrical power \cite{Kundur94}. These grids predominantly operate on alternating currents and can be considered networks of coupled rotating machines, with phase angle differences between connected machines determining the power flow between them \cite{dorfler_synchronization_2014,filatrella_analysis_2008,nishikawa_comparative_2015}. Practically, grid operators only allow for very small deviations of machine frequencies from the prescribed grid frequency (which is typically either 50Hz or 60Hz), as well as for only minor phase differences of connected oscillators in synchronized regimes. This guarantees efficient machine operations and a stable power flow as otherwise, rotating machines can be damaged, transmission lines trip and cascading failures be triggered \cite{Kundur94}. Knowing a grid's phase-locked solutions yields its stationary power flow; this helps identifying vulnerable transmission lines along which the power flow may exceed critical limits (see also \cite{witthaut_braesss_2012}), preventing line tripping and mitigating the risk of cascading failures \cite{Kundur94}. This enables a robust grid design and a cost-effective modification of power grids \cite{pinto_optimal_2015}. While voltage instabilities have been taken into account by some models \cite{schmietendorf_self-organized_2014}, one can assume constant voltages in high-voltage grids \cite{filatrella_analysis_2008, nishikawa_comparative_2015}. In the vicinity of the synchronized regime, the AC power flow can then be modeled with second-order Kuramoto-type equations \cite{filatrella_analysis_2008,nishikawa_comparative_2015}, with steady states and their linear stability still captured by the original Eq.~\ref{e:kuramoto} \cite{dorfler_synchronization_2013,manik_supply_2014}. Here, we want to investigate the impact of grid topology on frequency synchronization in Eq.~\ref{e:kuramoto}, in particular on the form of locked oscillator phases. To this end, it is legitimate to simplify machine dynamics and assume that in the grid, all generators inject the same power and all consumers draw the same power, respectively. This is reflected by a bipolar distribution of natural frequencies, partitioning the oscillator ensemble into  generators with a single positive natural frequency and consumers with a single negative natural frequency \cite{buzna_synchronization_2009,witthaut_braesss_2012,rohden_impact_2014}. 

For many collective processes, the influence of their topological background on system dynamics has been thoroughly investigated \cite{dorogovtsev_critical_2008}, including the aforementioned partial synchronization regime for coupled oscillators \cite{acebron_kuramoto_2005,arenas_synchronization_2008,dorfler_synchronization_2014,rodrigues_kuramoto_2016}. Yet in the case of frequency synchronization in oscillator networks, results on the effect of network structure are few and mostly concern the critical threshold \citep{dorfler_synchronization_2013,motter_spontaneous_2013,verwoerd_global_2008,rohden_impact_2014,buzna_synchronization_2009}. Concerning another crucial characteristic of synchronized regimes - an analytic expression for locked phases - we find two recent contributions to be the most relevant. In \cite{wang_approximate_2015}, locked phases are approximately calculated for a fully connected network with arbitrary distributions of natural frequencies. However, synchronized regimes just above critical thresholds are not captured for which locked phase differences exceed $\pi/2$.  In the collective-coordinate approach of \cite{gottwald_model_2015}, a functional form of locked phases is imposed through an educated guess, reducing the complexity of Eq.~\ref{e:kuramoto}. This allows for analytically tractable approximate evolution equations for the system whose steady state fully determine locked phases (see \cite{pinto_optimal_2015} for further applications). This approach can accommodate different coupling structures, yet requires the functional form of locked phases to be known \emph{a priori}. To the best of our knowledge, there exists no analytic approach to date that, for power grid-like oscillator networks, yields locked phases other than for quite restrictive assumptions on graph topology, correlations in natural frequencies or the form of locked phases.

Here, we want to fill this gap with an analytic framework that allows to systematically explore the influence of topology on frequency synchronization in power-grid like oscillator networks. To characterize connection structures, we consider two simple local measures: an oscillator's neighborhood size (its \emph{degree}) and its neighborhood composition with respect to natural frequencies. This enables us to describe locked phases for the broad class of configuration models: these are networks with an arbitrary imposed degree distribution, i.e., a probability distribution of oscillator degrees, while otherwise featuring random connections. We focus on the configuration model as it is usually taken as the starting point for investigating the effect of network heterogeneity on collective dynamics \cite{dorogovtsev_critical_2008,marceau_adaptive_2010,restrepo_onset_2005}. For the locked phases in this broad class of grid topologies, we determine the linear stability as well as identify a shared functional form and universal monotonic behavior. Note that a degree distribution does not uniquely define a network, but rather a network \emph{ensemble} comprised of all adjacency matrices whose row sums obey the given distribution. Consequently, our approach makes statements about network ensembles, predicting that all members of a given ensemble behave similarly. This contrasts with the focus on single network realizations taken in previous literature on frequency synchronization. In addition to departing from the regular-lattice assumption in \cite{buzna_synchronization_2009} and the fully connected setup in \cite{wang_approximate_2015}, we also differ from these contributions by allowing for phase differences to exceed $\pi/2$ at weakly supercritical coupling. Furthermore, we generalize \cite{buzna_synchronization_2009} through allowing for an \emph{asymmetric} bipolar distribution of \emph{randomly} assigned natural frequencies. This is a more realistic setting to investigate stable operations of power grids, and can be easily extended to include correlations in natural frequencies.

Our working hypothesis is that in correlation-free networks described by the configuration model, an oscillator’s locked phase is shaped by its neighborhood (along with its own natural frequency). This hypothesis is formalized in Sec.~\ref{s:mf} through a mean-field ansatz inspired by \cite{restrepo_onset_2005} and resembling the active-neighborhood approach previously used in the context of disease spreading \cite{marceau_adaptive_2010}. In the central Sec.~\ref{s:param}, we use the ansatz to obtain a parametrized description of the effect of connection structure on frequency synchronization in oscillator networks. In Sec.~\ref{s:self}, we present a simple self-consistent calculation of the respective parameter. We assess results in Sec.~\ref{s:comparison} for regular random graphs where links are randomly distributed under the constraint that all oscillators have the same degree, and for random graphs without aforementioned constraint. We summarize and give an outlook in Sec.~\ref{s:summary}, while technical details can be found in the appendix.
\section{Mean-field equations}\label{s:mf}
\begin{figure}[ht]
  \centering
    \includegraphics[width=0.9\textwidth]{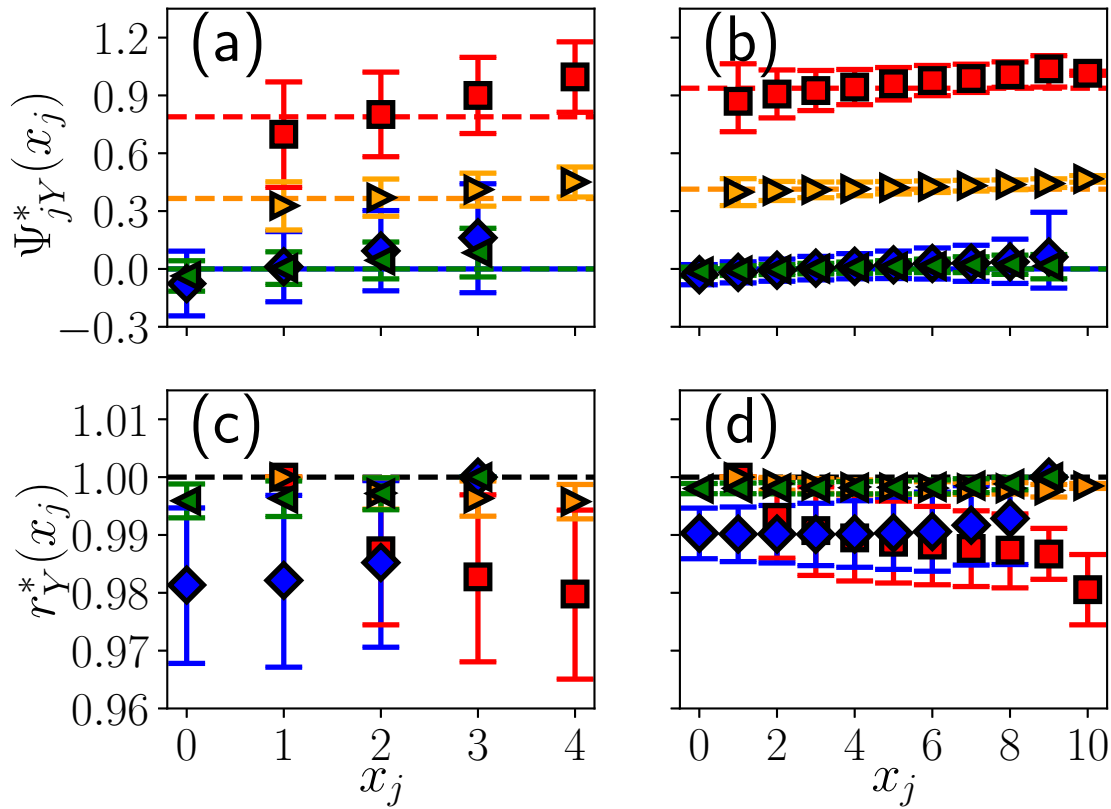}
  \caption{(Color online) Accuracy of mean-field assumptions with respect to numerical integration of Eq.~\ref{e:kuramoto} on regular random graphs. Phase-locked generator neighborhoods  $x_j$ with degree $k=4$ [(a), (c)] and $k=10$ [(b), (d)]. For $\lambda$ just above (well above) critical threshold, red squares (orange right triangles) depict averaged $\Psi^*_{jG}(x_j)\equiv \Psi^*_{jG}(k,x_j)$ [(a)-(b)] or $r^*_G(x_j)\equiv r^*_G(k,x_j)\equiv r^*_{jG}/x_j$ [(c)-(d)], blue diamonds (green left triangles) represent averaged $\Psi^*_{jC}(x_j)\equiv\Psi^*_{jC}(k,x_j)$ [(a)-(b)] or $r^*_C(x_j)\equiv r^*_C(k,x_j)\equiv r^*_{jC}/(k-x_j)$ [(c)-(d)]. Averages from one network realization with $N=10^6$, $g=0.3$ and all oscillators sampled. Error bars give data-point range between 16th and 84th percentile of respective distribution. (a) and (c): $\lambda=0.75$ and $\lambda=2$ ($\lambda^*\approx 0.69$). (b) and (d):  $\lambda=0.2$ and $\lambda=1$ ($\lambda^*\approx 0.17$). (a)-(b): dashed lines are mean phases of  respective neighborhood type ($\Psi_C\equiv 0$ for both coupling strengths). (c)-(d): black dashed line corresponds to perfect neighborhood phase coherence of 1. All phases given in radians.}\label{f:mf}
\end{figure}

In this section, we lay out the principal assumptions and practices used in the rest of this work. To model synchronization in power grids, we assume - for $g\in(0,1/2]$ -  an integer number $g N$ of generators with $\omega_j=1$ and $j\in[1,g N]$, as well as an integer number $(1-g)N$ of consumers with $\omega_j=-g/(1-g)$ and $j\in[g N+1,N]$. It follows that the system frequency in Eq.~\ref{e:kuramoto} is zero, and steady states are identified with frequency synchronization in the system. Moreover, we account for the observation that real-world power grids usually contain more consumers than generators. It is easy to check that the chosen natural frequencies and range of \(g\) capture all steady states of Eq.~\ref{e:kuramoto} and their linear stability. In the following, we mainly choose $g=0.3$ in numerical investigations, as this is a realistic choice for the fraction of generators in power grids \cite{nishikawa_comparative_2015}; yet theoretical considerations hold for all $g$. Furthermore, all oscillator phases are confined to $[0,2\pi)$, and all quantities pertaining to generators and consumers have subscripts "G" and "C", respectively. In the spirit of \cite{restrepo_onset_2005}, Eq.~\ref{e:kuramoto} can then be rewritten as 
\begin{equation}\label{e:kuramoto2}
\dot{\theta}_j=\omega_j+\lambda\!\left[r_{jG} \sin{(\Psi_{jG}-\theta_j)}+r_{jC} \sin{(\Psi_{jC}-\theta_j)}\right]
\end{equation}
with $r_{jG} e^{i\Psi_{jG}}\equiv\sum_{n=1}^{g N} A_{jn} e^{i\theta_n}$ and $r_{jC} e^{i\Psi_{jC}}\equiv\sum_{n=g N+1}^N A_{jn} e^{i\theta_n}$, multiplying both sides with $e^{-i\theta_j}$ in each case before equating the imaginary parts \cite{kuramoto_chemical_1984}. Hence the sum in Eq.~\ref{e:kuramoto} is split into two simple terms quantifying the coupling of oscillator $j$ to the group of adjacent generators and consumers, respectively. This comes at the cost of introducing four new variables per oscillator - one coupling amplitude $r_{jY}$ and one neighborhood phase $\Psi_{jY}$ for each neighborhood type $Y\in\{G,C\}$.  Thus, in order to proceed analytically, we make the following four simplifying assumptions:

\textbf{(a1)} \emph{We assume that the neighborhood phases $\Psi_{jG}$ and $\Psi_{jC}$ of any oscillator $j$ only depend on $j$'s natural frequency.} This presupposes that the pull of $j$'s phase on neighboring oscillators' phases is either (1) negligible, i.e., for sufficiently large neighborhoods or (2) dominated by $\omega_j$, implying that $\theta_j$ is largely determined by $\omega_j$. Hence for generators, we introduce $\Psi_{G}$, the mean phase of neighboring generators and $\Psi_{C}$, the mean phase of neighboring consumers [Figs.~\ref{f:mf}(a)-(b)]. In the same spirit, for consumers, we introduce $\hat{\Psi}_{G}$ as the mean phase of their neighborhood of generators and $\hat{\Psi}_{C}$ as the mean phase of their neighborhood of consumers. 

\textbf{(a2)} \emph{For generator $j$, we approximate the coupling amplitudes $r_{jG}$ and $r_{jC}$ as $r_{jG} =r_G  x_j$ and $r_{jC}=r_C (k_j-x_j)$, where $x_j$ and $k_j$ are the number of generator and the total number of neighbors of $j$, respectively.} This approximation supposes that there is a global mean field that faithfully reflects oscillator dynamics, and that an oscillator's coupling to it is proportional to the oscillator's local connectivity \cite{restrepo_onset_2005}. For well-connected, homogeneous graphs, this is the case \cite{rodrigues_kuramoto_2016,restrepo_onset_2005} [Figs.~\ref{f:mf}(c)-(d)]. The proportionality factors $r_G$ and $r_C$ can be understood as the mean phase coherences of the generator and consumer portion of generator neighborhoods, respectively. Analogously, $\hat{r}_G$ and $\hat{r}_C$ can be defined, describing mean phase coherences in the two neighborhood types of consumers.

To compute for instance $r_G$ and $\Psi_G$, first consider that with (a1) and (a2), $r_G e^{i\Psi_{G}}=\sum_{j=1}^{g N} r_{jG} e^{i\Psi_{jG}}/\sum_{j=1}^{g N} x_j$. Hence $r_G$ not only estimates the phase coherence of generator neighbors of a single given generator, but also measures the coherence of generator neighborhood phases $\Psi_{jG}$ across the whole generator ensemble. It therefore assesses the validity of assumption (a1), and does so the more accurately the better (a1)-(a2) are fulfilled. With $\sum_{j=1}^{g  N} A_{jn} e^{i\theta_n}=x_n e^{i\theta_n}$, it follows that $r_G e^{i\Psi_G}=\sum_{j=1}^{g  N} x_j e^{i\theta_j}/\sum_{j=1}^{g  N} x_j$.  Evidently $r_G\in[0,1]$ without loss of generality, so that a value of $r_G$ close to $1$ indicates the goodness of (a1)-(a2), as exploited further below in assumption (a3). Similar expressions hold for $r_C e^{i\Psi_C}$, $\hat{r}_G e^{i\hat{\Psi}_G}$ and $\hat{r}_C e^{i\hat{\Psi}_C}$, so that the coupling in Eq.~\ref{e:kuramoto2} of an oscillator $j$ to the mean field just depends on $k_j$, $x_j$ and $\omega_j$. Consequently, all oscillators with the same natural frequency and neighborhood can be considered equivalent and described by a phase $\theta_Y(k_j,x_j)$ with $Y\in\{G,C\}$  [cf. Figs.~\ref{f:rrg}(c)-(f) and Figs.~\ref{f:erg}(c)-(d)]. This is reminiscent of the active-neighborhood approach  presented in \cite{marceau_adaptive_2010} in the context of disease spreading, which captures configuration-model topologies characterized by the degree distribution $P(k)$ and its first moment $\langle k \rangle$. As frequency synchronization in a network presupposes the network to be connected, we set $P(0)=0$ in the following. Assuming random mixing of natural frequencies, generators and consumers follow the same binomial distribution in $x$ and the same distribution $P(k)$ in $k$. Hence, with $w_g(k,x)\equiv P(k) x \binom{k}{x}g^x(1-g)^{k-x}$ and $Y\in\{G,C\}$, one obtains
\begin{eqnarray}
r_{Y} e^{i\Psi_{Y}}&=&\sum_{k=1}^\infty \sum_{x=0}^k \frac{w_g(k,x)}{g \langle k\rangle}e^{i\theta_{Y}(k,x)}\label{e:ordG}\\
\hat{r}_{Y} e^{i\hat{\Psi}_{Y}}&=&\sum_{k=1}^\infty\sum_{x=0}^k \frac{w_{1-g}(k,k-x)}{(1-g)\langle k\rangle}  e^{i\theta_{Y}(k,x)}\label{e:ordC} \,.
\end{eqnarray}
In Eqs.~\ref{e:ordG}-\ref{e:ordC}, the time-dependent mean-field - initially defined through averaging over $r_{jY}  e^{i\Psi_{jY}}$, i.e., over oscillator neighborhoods - is expressed through weighted averages over $e^{i\theta_{Y}(k,x)}$, i.e., over oscillators. Note that in these weighted averages, oscillator classes contribute proportionally to their abundance and to the size of their relevant neighborhood type. This is in contrast to the classical definition of phase coherences that does not include neighborhood sizes in the weighting \cite{kuramoto_chemical_1984}. In the following, one can fix the value of each mean neighborhood phase coherence through a third assumption:

\textbf{(a3)} \emph{We set $r_{Y}= 1$ and $\hat{r}_{Y}= 1$, $Y\in\{G,C\}$.} As laid out above, this is consequential if one is convinced of the goodness of (a1)-(a2). More support for this assumption is given by the fact that the phase coherence of oscillators in a shared neighborhood is generally larger than this of oscillators of same type, but randomly picked in the network, since common neighbors are at most two links apart. As an increased intra-neighborhood phase coherence yields an increased inter-neighborhood phase coherence as quantified by $r_{Y}$ and $\hat{r}_{Y}$, (a3) follows [Figs.~\ref{f:mf}(c)-(d)].

With (a1)-(a3), Eq.~\ref{e:kuramoto2} can be written as
\begin{eqnarray}
\dot{\theta}_G(k,x)&=&1+\lambda x\sin{[\Psi_{G}-\theta_G(k,x)]}+\lambda(k-x)\sin{[\Psi_{C}-\theta_G(k,x)]}\label{e:mfG}\\
\dot{\theta}_C(k,x)&=&-g(1-g)^{-1}+\lambda x \sin{[\hat{\Psi}_{G}-\theta_C(k,x)]}+\lambda(k-x)\sin{[\hat{\Psi}_{C}-\theta_C(k,x)]}\label{e:mfC} \,.
\end{eqnarray}
Note that all information about network topology and frequency mixing is contained in Eqs.~\ref{e:ordG}-\ref{e:ordC} that define the mean field, while the evolution Eqs.~\ref{e:mfG}-\ref{e:mfC} for oscillator phases are unaffected otherwise. The latter's synchronized solutions rotate with frequency
\begin{equation}\label{e:freq}
\omega_S=-\lambda\langle k\rangle  g (1-g)(\hat{r}_G-r_C) \sin{(\hat{\Psi}_G-\Psi_C)}
\end{equation}
whose absolute value is a model-intrinsic measure of the quality of approximation (a3) (Sec.~\ref{s:rotation}). The rotation can be neglected through the follow-up assumption

\textbf{(a4)}
\emph{The form and linear stability of synchronized solutions of Eqs.~\ref{e:ordG}-\ref{e:mfC} is similar to the form and linear stability of the steady states of Eqs.~\ref{e:ordG}-\ref{e:mfC} with an arbitrarily chosen fixed phase.}

The validity of assumptions (a1)-(a3) for Eq.~\ref{e:kuramoto} is then illustrated in Figs.~\ref{f:mf}(a)-(d). There, we show generator neighborhoods in phase-locked states of regular random graphs. This is to specifically analyze how the mean field copes with heterogeneity in the neighborhood composition of oscillators. As expected, (a1)-(a3) become more accurate for larger degrees and coupling strengths. This is corroborated both by increasingly accurate averages and by decreasing variances for each oscillator class. Note that (a1)-(a3) already hold slightly above the respective critical threshold (red squares and blue diamonds), but also in Figs.~\ref{f:mf}(a)-(b) the slight dependence of mean neighborhood phases on the neighborhood composition. The latter implies a dependence of an oscillator's phase on its neighborhood type $x$ that is investigated further below [cf. Figs.~\ref{f:rrg}(c)-(f) and Figs.~\ref{f:erg}(a)-(b)]. Similar observations hold for consumer neighborhoods, thus additionally confirming (a4) via Eq.~\ref{e:freq} (not shown). Furthermore, numerical integration confirms very similar critical thresholds and values of locked phases differences for Eqs.~\ref{e:ordG}-\ref{e:mfC} with and without assumption (a4) (not shown). Thus with (a4), one can approximate synchronized solutions of the mean-field Eqs.~\ref{e:ordG}-\ref{e:mfC} with  its steady-state solutions for  slightly altered natural frequencies. This makes the system amenable to analytic treatment, as laid out in the following section.
\section{Parametrized locked generator phases}\label{s:param}
In the following, we analytically identify characteristics that hold for synchronized regimes on all configuration-model topologies. As a consequence of assumptions (a1)-(a3), the Kuramoto model in Eq.~\ref{e:kuramoto} is well-approximated by two sets of neighborhood-class Eqs.~\ref{e:ordG}-\ref{e:mfC} - one set for each oscillator type. These two sets can be disentangled through assumption (a4) and an appropriate choice of coordinates: with $\Psi_C\equiv 0$, $\Psi\equiv \Psi_G$ and $\theta_\Psi(k,x)\equiv\theta_G(k,x)$, Eqs.~\ref{e:ordG} and \ref{e:mfG} become
\begin{eqnarray}
\dot{\theta}_\Psi(k,x)&=&1-\lambda \{x\sin{\left[\theta_\Psi(k,x)-\Psi\right]}+(k-x)\sin{\left[\theta_\Psi(k,x)\right]}\}\label{e:mfG2M}\\
\Psi&=&\arg\left[\sum_{k=k_m}^{k_M}\sum_{x=0}^k{w_g(k,x)e^{i\theta_\Psi(k,x)}}\right]\label{e:ordG2M}\, .
\end{eqnarray}
Consequently, generator phases are just coupled to a single mean-field variable $\Psi$. This system co-evolves self-containedly and is analyzed in the following; consumer dynamics can be decoupled and dealt with analogously (see further below).

A necessary condition for global phase-locking is that all generator classes $(k,x)$ phase-lock, which is surely the case if fixed points of Eqs.~\ref{e:mfG2M}-\ref{e:ordG2M} are linearly stable for all $(k,x)$. The complexity of this task can be reduced by first considering $\Psi$ a constant parameter in Eq.~\ref{e:mfG2M} that is self-consistently computed via Eq.~\ref{e:ordG2M}. This decouples the dynamics of each generator class not just from consumers, but also from other generator classes. The above phase-locking condition then translates into searching for values of $\Psi$ which (1) yield a linearly stable fixed point in Eq.~\ref{e:mfG2M} for all $(k,x)$ (2) fulfill Eq.~\ref{e:ordG2M} and (3) yield a linearly stable fixed point in co-evolving Eqs.~\ref{e:mfG2M}-\ref{e:ordG2M}. The solution to the generator phase-locking problem can thus be divided into three parts: describing the steady state of Eq.~\ref{e:mfG2M} parameterized by a fixed $\Psi$ (Appendix~\ref{s:closedForm}) while characterizing some of its general properties (Appendix~\ref{s:properties}), self-consistently computing $\Psi$ with Eq.~\ref{e:ordG2M} (Appendix~\ref{s:selfConsistent}), with discarding unstable solutions (Appendices~\ref{s:closedForm}-\ref{s:selfConsistent}) along the way .

\begin{figure}[ht]
  \centering
    \includegraphics[width=0.9\textwidth]{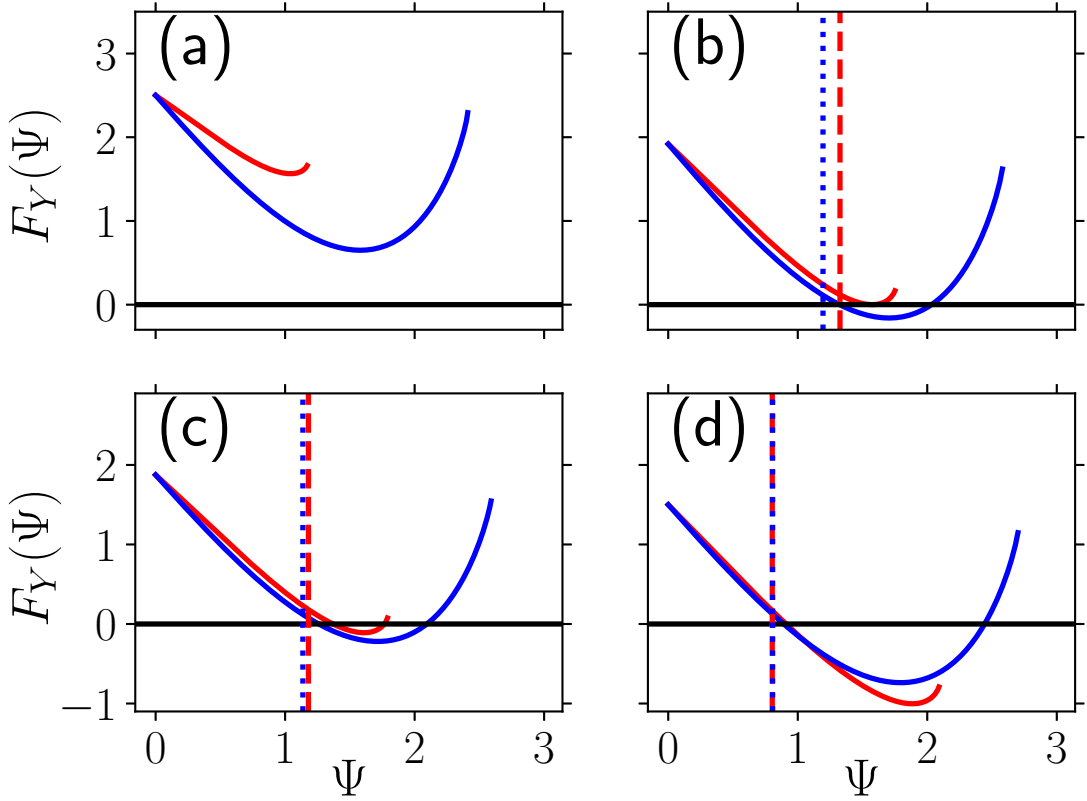}
  \caption{(Color online) Self-consistent computation of $\Psi$ in regular random graph with $k=10$ and $g=0.3$. $F_{G}(\Psi)$ (Eq.~\ref{e:psi2M}, red solid line) and $F_{C}(\Psi)$ (Eq.~\ref{e:Fc}, blue solid line) for different coupling strengths. Intersections with black solid line yield self-consistent $\Psi^*_G$ and $\hat{\psi}^*_C$, respectively. Vertical red dashed (blue dotted) lines give values for steady-state $\hat{\Psi}^*_G$ ($\psi^*_C$ ) at linearly stable solution for $\Psi^*_G$ ($\hat{\psi}^*_C$). Note that usually $\Psi^*_G\neq \hat{\Psi}^*_G$ and $\psi^*_C\neq \hat{\psi}^*_C$, and that $F_G(\Psi)$ as well as $F_C(\Psi)$ are restricted to intervals in Eq.~\ref{e:intervalM} and consumer counterparts.  (a) $\lambda=0.12$. (b) $\lambda=\lambda^*_G\approx 0.156$. (c) $\lambda=0.16$. (d) $\lambda=0.2$. All phases given in radians.}\label{f:self}
\end{figure}

Setting $\Psi$ constant without specifying its value, vital properties of locked generator phases can already be analytically derived. Appendix~\ref{s:closedForm} identifies and discards linearly unstable phase-locked states in Eq.~\ref{e:mfG2M}, and calculates the closed-form expressions for linearly stable locked generator phases as
\begin{eqnarray}
\sin{\left[\theta^*_\Psi(k,x)\right]}&=&\frac{\lambda^{-1}[k-(1-\cos{\Psi})x]+x\sin{\Psi} \sqrt{k^2-2x(k-x)(1-\cos{\Psi})-\lambda^{-2}}}{k^2-2x(k-x)(1-\cos{\Psi})}\label{e:solsineM}\\
\cos{[\theta^*_\Psi(k,x)]}&=&\frac{\sin{[\theta^*_\Psi(k,x)]}[k-x(1-\cos{\Psi})]-\lambda^{-1}}{x\sin{\Psi}}\,,\label{e:solcosineM}
\end{eqnarray}
with $\cos{[\theta^*_\Psi(k,0)]}=\sqrt{1-(\lambda k)^{-2}}$. Appendix~\ref{s:selfConsistent} simplifies the self-consistent computation of $\Psi$ in Eq.~\ref{e:ordG2M} to 
\begin{equation}\label{e:psi2M}
0=\sum_{k=k_m}^{k_M}\sum_{x=0}^k w_g(k,x) \sin{[\theta^*_{-\Psi}(k,k-x)]}\equiv F_G(\Psi)
\end{equation}
with
\begin{equation}\label{e:intervalM}
\Psi \in\begin{cases}
\left[0,\arccos{\left(2[\lambda k_m]^{-2}-1\right)}\right]\\
\left[0,\arccos{\left(\max\{-1,2(\lambda k_m)^{-2}-1-2\frac{1-(\lambda k_m)^{-2}}{k_m^2-1}\}\right)}\right] 
\end{cases}
\end{equation}
in case the smallest realized generator degree $k_m$ in the network is even- or odd-valued, respectively, and $\Psi\in(0,\pi)$ for odd-valued $k_m$ being reasonably asserted. Appendix~\ref{s:properties} identifies universal monotonic behavior of locked phases; it shows that on the above interval,
\begin{equation}\label{e:monotonyM}
\theta^*_\Psi(k,x+1)\geq \theta^*_\Psi(k,x)
\end{equation}
and
\begin{equation}\label{e:slopeM}
\theta^*_\Psi(k',k'x/k)\leq \theta^*_\Psi(k,x)
\end{equation}
for integer $k'>k$ and $k'x/k$ as well as
\begin{equation}\label{e:spreadM}
\theta^*_\Psi(k,x) \in[\arcsin{(\lambda k)^{-1}},\Psi+\arcsin{(\lambda k)^{-1}}]
\end{equation}
for stable locked generator phases, regardless of chosen degree distribution $P(k)$. Lastly, the ensemble phase spread - the maximum difference between two locked generator phases - is shown to be exactly $\Psi$, so that its upper bound is trivially given by Eq.~\ref{e:intervalM}. This upper bound becomes the tighter the closer the coupling strength is to the critical threshold. This is because computed stable steady-state $\Psi$ in Eq.~\ref{e:psi2M} are the closer to the upper bound of the search interval in Eq.~\ref{e:intervalM} the smaller the coupling strength [Figs.~\ref{f:self}(b)-(d)]. Equations~\ref{e:solsineM}-\ref{e:solcosineM} and \ref{e:monotonyM}-\ref{e:slopeM} are the central result of this work and are predicted to hold on any configuration-model topology, provided that assumptions (a1)-(a4) are met.

\section{Determining parameter and locked consumer phases}\label{s:self}
Here we self-consistently determine the stable steady-state $\Psi$ that parametrize derived expressions, and show how locked consumer phases are obtained from the synchronized generator ensemble. To this end, decoupled generator dynamics with a co-evolving mean-field $\Psi$ are captured in two steps: first, the parameter $\Psi$ is self-consistently computed through Eq.~\ref{e:psi2M} on the interval given by Eq.~\ref{e:intervalM}. Second, a linear stability analysis of co-evolving Eqs.~\ref{e:mfG2M}-\ref{e:ordG2M} is performed at the fixed points given by self-consistent $\Psi$ and the locked generator phases they parametrize in Eqs.~\ref{e:solsineM}-\ref{e:solcosineM} (cf. Appendix~\ref{s:selfConsistent}).

To illustrate the procedure, we choose in Figs.~\ref{f:self}(a)-(d) a regular random graph as the network topology. There, self-consistent values of $\Psi$ - the roots of $F_G(\Psi)$ (red solid line) - are computed. For sufficiently small $\lambda$, no self-consistent solution for $\Psi$ exists [Fig.~\ref{f:self}(a)] - there are no phase-locked generators. At $\lambda\equiv\lambda^*_G$, two solutions arise [Figs.~\ref{f:self}(b)-(c)]. The larger-valued solution is computed to be linearly unstable  and discarded. It disappears for sufficiently large $\lambda$ to leave only one (small-angle) solution for $\Psi$ that is always linearly stable [Fig.~\ref{f:self}(d)]. Consequently, $\lambda^*_G$ is the critical coupling strength in the decoupled generator ensemble, and generator phases can be stably locked in only one configuration. With $\Psi^*_G\equiv\Psi$ in the unique stable locked regime at hand, $\hat{\Psi}^*_G$ - the steady-state mean phase of generator neighbors of a consumer - is obtained by inserting Eqs.~\ref{e:solsineM}-\ref{e:solcosineM} into Eq.~\ref{e:ordC} [red dashed line in Figs.~\ref{f:self}(b)-(d)]. Consequently, analyzing the decoupled generator dynamics in co-evolving Eqs.~\ref{e:mfG2M}-\ref{e:ordG2M} delivers three things: first the generators' critical threshold $\lambda^*_G$, second the generators' linearly stable locked phases $\theta^*_G(k,x)$, third the steady-state mean neighborhood phases $\Psi^*_G$ and $\hat{\Psi}^*_G$. All phases are given in coordinates in which generator natural frequencies are positive and $\Psi_C\equiv 0$, the latter as imposed in Eqs.~\ref{e:ordG}-\ref{e:mfC} to decouple the two oscillator ensembles and obtain generator Eqs.~\ref{e:mfG2M}-\ref{e:ordG2M}. 

Consumer dynamics can be described in the same self-contained manner by setting $\hat{\Psi}_G\equiv 0$ and $\hat{\Psi}_C\equiv -\Psi$ in Eqs.~\ref{e:ordC} and \ref{e:mfC}. This is possible because the asymmetry of the two oscillator ensembles only lies in $g<1/2$ and in the resulting differing absolute value of their natural frequencies. Thus substituting $\lambda\rightarrow \lambda(1-g)/g$ in Eqs.~\ref{e:solsineM}-\ref{e:solcosineM} and \ref{e:intervalM}-\ref{e:spreadM} yields the respective quantities for the consumer ensemble if $x$ is now considered the number of \emph{consumer} neighbors. With this change of parameters and indices, all considerations in Appendices~\ref{s:closedForm}-\ref{s:properties} also apply to the decoupled consumer ensemble. The change in neighborhood indexing moreover leads to 
\begin{equation}\label{e:Fc}
F_C(\Psi)\equiv\sum_{k=k_m}^{k_M}\sum_{x=0}^k w_{1-g}(k,x) \sin{[\theta^*_{-\Psi}(k,k-x)]}
\end{equation}
(with $\lambda\rightarrow \lambda(1-g)/g$ in $\sin{[\theta^*_{-\Psi}(k,x)]}$, cf. Eq.~\ref{e:solsineM}) for the function  whose roots yield the self-consistent mean-field $\Psi$ for the consumer ensemble. Numerically, one finds for the consumer ensemble in the regular random graph that $\lambda^*_C<\lambda^*_G$ for all considered $g<1/2$ and $k$ [see blue solid line in Figs.~\ref{f:self}(a)-(d)], so that the critical threshold of the full system is given by $\lambda^*=\lambda^*_G$. This can be qualitatively understood by considering that the lower bound on the critical threshold for generators is larger than that for consumers by a factor of $(1-g)/g$ (cf. Eq.~\ref{e:synchEvenA}). Upon computation of the self-consistent stable $\Psi\equiv \hat{\psi}^*_{C}$ - the steady-state mean phase of consumer neighbors of consumers -  all consumer phases are given in a reference frame in which their mean generator-neighborhood phase $\hat{\psi}^*_{G}$ is zero and their natural frequency is of positive value. For the stable steady state, the locked phases $\Theta^*_C(k,x)$ and $\psi^*_C$ [blue dashed line in Figs.~\ref{f:self}(b)-(d) for a regular random graph] are then computed analogously to the generator ensemble. 

To relate locked generator phases [$\theta^*_G(k,x),\Psi^*_G,\Psi^*_C,\hat{\Psi}^*_G$] and locked consumer phases [$\Theta^*_C(k,x),\hat{\psi}^*_G,\hat{\psi}^*_C,\psi^*_C$], one should express them in common coordinates. This can be done by (1) substituting $x\rightarrow k-x$ in $\Theta^*_C(x)$ (2) changing the sign of all consumer phases (3) additionally rotating all consumer phases by an angle $\rho$ so that $-\hat{\psi}^*_G+\rho=\hat{\Psi}^*_G$ and $-\psi^*_C+\rho=\Psi^*_C$. Step (1) expresses consumer classes based on generator neighbors, while step (2) accounts for the differing signs of the two natural frequencies. Note also that steps (1) and (2) leave Eq.~\ref{e:monotonyM} valid for the locked consumer phases $\theta^*_C(k,x)$ in generator coordinates, whereas Eq.~\ref{e:slopeM} becomes
\begin{equation}\label{e:slopeMC}
\theta^*_C(k',k'x/k)\geq \theta^*_C(k,x)
\end{equation}
for $k'>k$ and integer $k'x/k$. Finally, step (3) relates the two sets of locked phases by demanding that computing the same phase differently should yield (approximately) the same result. As a result, the rotation angle is over-determined as $\rho=\psi^*_C$ and $\rho=\hat{\Psi}^*_G$, and the extend to which both equations yield similar $\rho$ gives a measure of the quality of approximations (a1)-(a4) that led to Eqs.~\ref{e:mfG2M}-\ref{e:ordG2M}. Figures \ref{f:self}(b)-(d) show a good match already for weak supercritical coupling strengths in the case of the regular random graph, complementing Figs.~\ref{f:mf}(a)-(d) in supporting the validity of our mean-field assumptions for this grid architecture. With the absence of bistability in both decoupled oscillator ensembles, one can also exclude for that type of regular random graphs the existence of bistable phase-locked regimes in the full mean-field description of coupled generator and consumer dynamics.
\section{Comparison with full system}\label{s:comparison}
\begin{figure}[ht]
  \centering
    \includegraphics[width=0.9\textwidth]{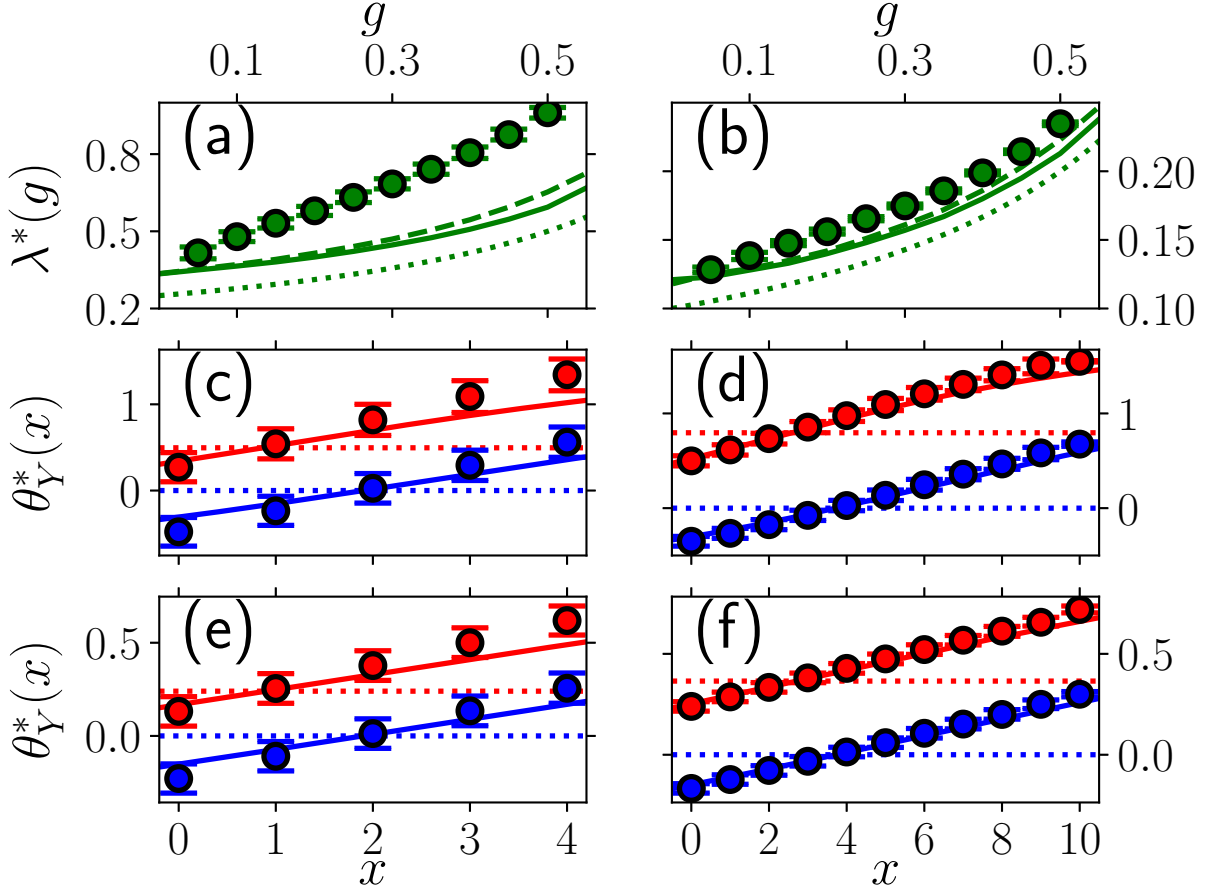}
  \caption{(Color online) Comparison of phase-locked states in mean-field approach and numerical integration of Eq.~\ref{e:kuramoto} on regular random graphs with $k=4$ [(a), (c), (e)] and $k=10$ [(b), (d), (f)]. Symbols: averages from full system in Eq.~\ref{e:kuramoto}. Solid lines: predictions from Eqs.~\ref{e:solsineM}-\ref{e:psi2M} and consumer counterparts. Dashed lines: predictions from Eqs.~\ref{e:kuramoto2} with only assumptions (a1)-(a2). Dotted lines: predictions from naive mean field. Error bars give  data-point range between 16th and 84th percentile of respective distribution. (a)-(b): critical thresholds for different $g$. Data  averaged over $10^3$ network realization with $10^4$ oscillators each. (c)-(f): locked phases $\theta^*_Y(x)\equiv \theta^*_Y(k,x)$ of generators (red) and consumers (blue) for $g=0.3$. Dashed lines coincide with solid lines. Data from one network realization with $10^6$ oscillators, all sampled. (c): $\lambda=0.75$. (d): $\lambda=0.2$. (e): $\lambda=1.5$. (f): $\lambda=0.4$.  All phases given in radians.}\label{f:rrg}
\end{figure}

With the mean-field framework fully laid out, we now compute critical thresholds and locked phases through Eqs.~\ref{e:solsineM}-\ref{e:psi2M} and their consumer counterparts. This is in turn compared with numerical integration of the full system in Eq.~\ref{e:kuramoto}. To illustrate the role of neighborhood composition in phase-locking, we first choose again  regular random graphs [sparse graphs in Figs.~\ref{f:rrg}(a), (c), (e), denser graphs in Figs.~\ref{f:rrg}(b), (d), (f)]. As in Figs.~\ref{f:mf}(a)-(d), we immediately notice decreasing intra-class variances for increasing graph connectivity and coupling strength. This confirms that our nearest-neighbor approach works best for well-connected graphs where longer-range correlations cause little variability of locked phases within the same oscillator class.

For the computation of the critical thresholds in Figs.~\ref{f:rrg}(a)-(b), we demand that $\sum_{j=1}^{N}\dot{\theta}^2_j/N\leq 10^{-12}$ after  $t=10^3$ for the network to be considered in a frequency-synchronized state. One observes that the closer $g$ is to $1/2$, the larger is the critical threshold. This is plausible, as then the difference in natural frequencies to bridge for synchronization increases, while the phase pull of the weaker generator ensemble also increases due to the ensemble's increasing size. Our framework (i) systematically underestimates $\lambda^{*}$; deviations grow with (ii) decreasing $k$ and (iii) increasing $g$. The reason for (i) is that our ansatz only distinguishes oscillators based on their natural frequency and neighborhood type, yielding a mean-field description for the phase-locking of each oscillator class. Yet the onset of frequency synchronization in real networks presupposes that also outlier oscillators phase-lock - oscillators whose network embedding (i.e., connections beyond their neighborhood) is particularly detrimental to their synchronization with respect to the mean-field description of their class. These outliers determine the onset of synchronization in real networks, retarding it with respect to our mean-field approach. This retardation is particularly pronounced for sparse connectivity, as there fluctuations in the topology and frequency composition of higher-order neighborhoods are more manifest in oscillator dynamics, thus explaining (ii). Finally, higher-order frequency-composition fluctuations with retarding effect are more pronounced the closer $g$ is to $1/2$, accounting for (iii). This is because, combinatorially speaking, there are more higher-order neighborhood configurations possible for more similar ensemble sizes. A higher number of configurations also means a higher chance of fluctuations with retarding effect on phase-locking. For the same reason, the retarding effect of outliers is more pronounced in topologies with broader degree distributions like the random graphs dealt with further below.

For increasing supercritical coupling strengths, the importance of said fluctuations diminishes [compare Figs.~\ref{f:rrg}(c) and (e) as well as Figs.~\ref{f:rrg}(d) and (f)]. This becomes apparent first through the increasingly accurate prediction of locked phases by our framework (i.e., by considering Eqs.~\ref{e:solsineM}-\ref{e:psi2M}, \ref{e:Fc} and \ref{e:ordG}-\ref{e:ordC} in that order), second in the full system through the decreasing variances of locked phases within one oscillator class. Moreover, plotting in Figs.~\ref{f:rrg}(a)-(f) the output of Eq.~\ref{e:kuramoto2} with only assumptions (a1)-(a2) yields very similar results to our final approach that also includes (a3)-(a4) [dashed lines in Figs.~\ref{f:rrg}(a)-(b), complete overlap with solid lines in Figs.~\ref{f:rrg}(c)-(f)]. This underlines the quality of (a3)-(a4) and indicates that (a1)-(a2) are the most significant source of deviation from the full system given by Eq.~\ref{e:kuramoto}. As predicted in Eq.~\ref{e:monotonyM}, locked phases increase monotonously with $x$, and, for fixed natural frequency, spread no farther apart than predicted by Eqs.~\ref{e:intervalM} and \ref{e:spreadM} and their consumer counterparts. We note that for coupling strengths just above the critical threshold, the ensemble and thus also the global phase spread can exceed $\pi/2$ [Figs.~\ref{f:rrg}(c)-(d)]. This is a phase-locked regime commonly not considered in literature where instead, authors build on a result from \cite{dorfler_synchronization_2013} that shows the linear stability of all steady states with global phase spread smaller than $\pi/2$. Our model is also in line with this result: observed unstable self-consistent solutions of Eq.~\ref{e:psi2M} for the regular random graph all possess global phase spreads larger than $\pi/2$.

\begin{figure}[ht]
  \centering
    \includegraphics[width=0.9\textwidth]{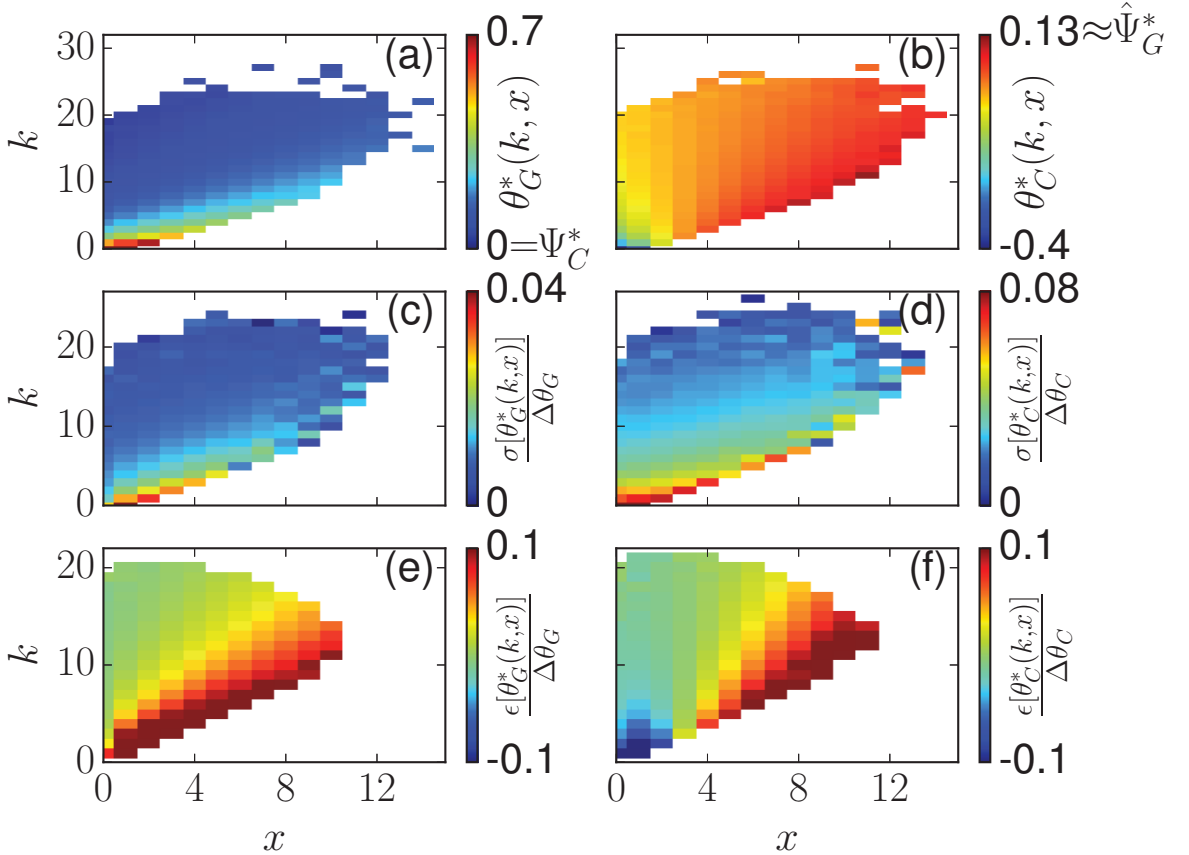}
  \caption{(Color online) Assessing accuracy and relevance of mean-field approach in Erd\H{o}s-R\'{e}nyi graphs. Locked generator phases $\theta^*_G(k,x)$ [(a), (c), (e)] and consumer phases $\theta^*_C(k,x)$ [(b), (d), (f)]. (a)-(b): locked phases in the full system. (c)-(d): Standard deviation for locked phases in full system, normalized by respective ensemble phase spread in full system. (e)-(f): error of mean-field prediction for locked phases, normalized by respective ensemble phase spread in full system. $\lambda=1.1$, $\langle k\rangle=10$, $g=0.3$. Data from full system for one network realization with $10^6$ oscillators, all sampled. All phases given in radians.}\label{f:erg}
\end{figure}

Furthermore, we observe in Appendix~\ref{s:comparison} that our framework is consistent with a more naive mean-field approximation, which in our setup moreover coincides with the collective-coordinate approach \cite{gottwald_model_2015} for regular random graphs. It yields $\theta^*_G(x)\equiv\Psi_G=\arcsin\left([\lambda k (1-g)]^{-1}\right)$ and $\theta*_C(x)\equiv \Phi_C\equiv 0$ above the critical threshold $\lambda^*= [ k(1-g)]^{-1}$. In Figs.~\ref{f:rrg}(a)-(f), the predictions of the naive mean-field model are plotted with dotted lines, showing that the inclusion of neighborhood heterogeneity in our more advanced approach yields a much better approximation of Eq.~\ref{e:kuramoto}.

The application of the mean-field approach to connected random (Erd\H{o}s-R\'{e}nyi) graphs with $P(k)=\langle k \rangle ^k  e^{-\langle k \rangle}/k!$ is illustrated in Figs.~\ref{f:erg}(a)-(f). For the chosen mean degree $\langle k\rangle=10$ and network size $N=10^6$, the minimum considered degree is $k_m=1$ and maximum degree $k_M=28$ (cf. Sec.~\ref{s:mf}), resulting in a cutoff for larger degrees [extended white area in Figs.~\ref{f:erg}(e)-(f), cf. Sec.~\ref{s:mf}]. With $k_m=1$ and Eq.~\ref{e:synchOddA}, a necessary condition for global phase-locking is $\lambda\geq 1$. Choosing $\lambda=1.1$, we find the full system to be in a weakly supercritical regime in which we test the validity of our mean-field approach.

We observe in Figs.~\ref{f:erg}(a)-(b) that the monotony of oscillator phases in the full system of Eq.~\ref{e:kuramoto} is correctly predicted by Eqs.~\ref{e:monotonyM}, \ref{e:slopeM} and \ref{e:slopeMC}. Furthermore, locked oscillator phases converge to the reference phase $\Psi_C\equiv 0$ for large degrees. This is because the larger an oscillator's neighborhood is, the stronger is its coupling to the rest of the network and the mean-field variable describing it. That also explains the convergence of locked consumer phases towards $\hat{\Psi}^*_G$ for increasing degrees [Fig.~\ref{f:erg}(b)], as this is the mean phase consumers couple to in their ensemble description. As in the case of regular random graphs, the overall phase spread exceeds $\pi/2$.  For the full system, the variances of locked phases within an oscillator class is very low for both oscillator ensembles [Figs.~\ref{f:erg}(c)-(d)], also for classes describing oscillators with lower degrees. This validates our mean-field approach of only considering oscillator neighborhoods when modeling phase-locking in configuration models. We normalize by the respective ensemble spread to obtain the relative magnitude of fluctuations. Lastly, Figs.~\ref{f:erg}(e)-(f) illustrate the differences of averaged locked phases in the full system and predicted locked phases, again normalized by the respective ensemble spread. They show that our framework predicts well the locked phases in the full system already for weak supercritical coupling strengths. As in Figs.~\ref{f:rrg}(c)-(f) for regular random graphs, we find a very good match for oscillator classes $(k,x$) for which $x\approx k g$, while predicted locked phases deviate the more the larger are deviations of $x$ from that mean number of generator neighbors.

\section{Summary and conclusions}\label{s:summary}
We present a mean-field approach to analytically assess how the phase-locking of randomly coupled oscillators is affected by their local connection structure. To that end, we consider asymmetric bipolar distributions of natural frequencies, modeling the most common asymptotic regime in electrical power grids. For those systems, we make two predictions, provided that the underlying mean-field assumptions are fulfilled reasonably well. Firstly, locked phases in all networks of the heterogeneous class of configuration models follow the same functional form and monotonic behavior. As the configuration model makes statements about entire network ensembles, we predict secondly that all members of a given ensemble have approximately the same locked phases and linear stability. This contrasts with previous contributions that each focus on a specific (and rather simple) grid topology. The first, \emph{inter-ensemble} prediction is substantiated numerically with two non-trivial standard topologies for two different network link densities and two different dynamical regimes each. The second, \emph{intra-ensemble} prediction becomes evident in the small standard deviations of ensemble statistics throughout this work. Both results do not depend on the specific shape of the bipolar distribution of natural frequencies, the latter of which is completely determined by the fraction of generators in the network. They show that while local connection structure does not govern the transition frequency synchronization \cite{dorfler_synchronization_2014}, it shapes the phase-locked state. The practical use of these results is that they help mitigate transmission line tripping and cascading failures.

The route to these main results is the following: Starting from a mean-field approach [Eqs.~\ref{e:ordG}-\ref{e:mfC} with assumption (a4)], generator and consumer ensembles are decoupled, each of which is analyzed separately and analogously (Eqs.~\ref{e:mfG2M}-\ref{e:ordG2M}). For each ensemble, unstable phase-locked regimes are detected and discarded, and stable locked phases are given closed-form parametrized expressions (Eqs.~\ref{e:solsineM}-\ref{e:solcosineM}). These expressions do not explicitly depend on imposed degree distributions; such global network information is instead contained in a mean-field parameter to which the considered oscillator ensemble couples (Eq.~\ref{e:ordG2M}). The search space for that parameter is analytically constrained (Eqs.~\ref{e:intervalM}), yielding simple statements about the locked phases' bounds (Eq.~\ref{e:spreadM}) and monotony (Eqs.~\ref{e:monotonyM}, \ref{e:slopeM} and \ref{e:slopeMC}) already without knowing the parameter's numerical value. 
For each ensemble, the mean-field parameter is self-consistently calculated through a simple expression (Eqs.~\ref{e:psi2M} and ~\ref{e:Fc}, respectively). Upon calculation, we find that above an ensemble-specific critical threshold, there exists a linearly stable steady state in each decoupled ensemble dynamics (Appendix~\ref{s:selfConsistent}). For coupled ensemble dynamics, i.e., the full mean-field model, the two ensemble-specific critical thresholds automatically yield the overall critical threshold. In order to actually couple ensemble dynamics, an overdetermined rotation angle finally yields the necessary coordinate transformation of locked consumer phases.

Apart from mentioned main results, we obtain further insights into the system: The discrepancy of the rotation angle's calculated values is a model-intrinsic measure for the accuracy of used mean-field assumptions, as is (the absolute value of) the system frequency of the mean-field equations (Eq.~\ref{e:freq}). Both measures indicate the validity of our mean-field approximations already for sparse and weak coupling. Additionally, we find that our approach is consistent with a simpler mean-field ansatz, lending further support our working hypothesis. We discover that for examined parameters and topologies, the framework rules out coexisting linearly stable phase-locked regimes that conform to mean-field assumptions (a1)-(a4). Furthermore, we find that for coupling strengths just above the critical threshold, the largest difference of locked phases can exceed $\pi/2$ in both the full system and its mean-field approximation - a regime commonly not considered in literature \cite{dorfler_synchronization_2014,wang_approximate_2015}, but captured by our approach. For sparsely connected or heterogeneous grids, critical thresholds are systematically underestimated; this is tied to the importance of outlier oscillators that do not obey mean-field assumptions.

In future work, one could explore a generalization of the functional form of locked phases to oscillator networks with more heterogeneous distributions of natural frequencies. Additionally, a more systematic investigation with our framework could reveal power-grid topologies with coexisting phase-locked regimes. While such multistability has been detected for single network realizations \cite{mehta_algebraic_2015,manik_multistability_2017}, our framework can potentially detect multistable regimes of a whole network ensemble given by the respective configuration model. Furthermore, by adjusting the binomial weights in Eqs.~\ref{e:ordG}-\ref{e:ordC}, our assumption of random frequency mixing could be relaxed to account for frequency correlations as in \cite{skardal_frequency_2015}. Finally, by yielding the functional form of locked phases \emph{a posteriori}, our proposed approach could moreover tie dimension-reduction approaches such as in \cite{gottwald_model_2015,ott_low_2008,martens_exact_2009} to intuitive mean-field assumptions.

\begin{acknowledgments}
The authors would like to thank F. Daviaud, F. Suard, M. Velay and M. Vinyals for useful discussions. Funding by CEA under the NTE NESTOR grant is gratefully acknowledged. Additionally, S.W. acknowledges funding of project 330 in the Enhanced Eurotalents program.
\end{acknowledgments}
\appendix
\section{System frequency of mean-field equations}\label{s:rotation}
Multiplying both sides of Eq.~\ref{e:mfG} [of Eq.~\ref{e:mfC}] with $g  w_g(k,x)/x$ [with $(1-g)  w_g(k,x)/x$], summing over $x$ and $k$ and adding both equations, one obtains with Eqs.~\ref{e:ordG}-\ref{e:ordC} 
\begin{equation*}
\omega_S=-\lambda\langle k\rangle  g (1-g)(\hat{r}_G-r_C) \sin{(\hat{\Psi}_G-\Psi_C)}
\end{equation*}
for $\omega_S\equiv\sum_{k=1}^\infty \sum_{x=0}^k\{[g\dot{\theta}_G(k,x)+(1-g)\dot{\theta}_C(k,x)]w_g(k,x)/x\}$ as the system frequency of the mean-field Eqs.~\ref{e:ordG}-\ref{e:mfC}. ote that in our reference frame, the system frequency in the full Eq.~\ref{e:kuramoto} is in contrast identical to zero. Here $r_C$ and $\hat{r}_G$ are not set to $1$, but computed on the fly from Eqs.~\ref{e:ordG}-\ref{e:ordC}. Hence any inaccuracy in assumption (a3) translates into a nonzero system frequency of the mean-field equations. Concurrently, it is easy to show along the same lines that Eq.~\ref{e:kuramoto2} with just assumptions (a1)-(a2) (i.e., allowing for variable neighborhood phase coherences) has zero system frequency at any time $t$. From the approximative character of (a3) follows that if the numerical integration of Eqs.~\ref{e:ordG}-\ref{e:mfC} for sufficiently large $\lambda$ does yield a synchronous regime (i.e., locked phase differences), it is generally not in steady state - despite the initial rescaling of natural frequencies to enter a co-rotating reference frame. Instead, all phases then synchronously rotate with the system frequency $\omega_S$. If $\omega_S$ is subtracted from both sides of Eqs.~\ref{e:mfG}-\ref{e:mfC}, synchronization is reflected by a steady state of oscillators with natural frequencies shifted by $-\omega_S$ and all phases at time $t$ rotated by $- \omega_S t$. 

Obviously, $|\omega_S|$ is the smaller the smaller $(\hat{r}_G-r_C)$, i.e., the better assumption (a3) is fulfilled. It vanishes for $g\rightarrow 0$ and $g=1/2$, the latter due to the resulting symmetry in Eqs.~\ref{e:ordG}-\ref{e:mfC} which yields $\hat{r}_G=r_C$. Yet $|\ \omega_S|$ also decreases with increasing $\lambda$ and $k$: both yields decreasing $(\hat{\Psi}_G-\Psi_C)$ and a higher accuracy of the mean field, i.e., also decreasing $|\hat{r}_G-r_C|$. Consequently, $\omega_S$ in phase-locked regimes is largest for slightly supercritical $\lambda$ and small $k$, usually attaining values of less than one percent of generator and consumer frequencies. This small value justifies equating $\hat{r}_G$ and $r_C$ in assumption (a3) and thus also setting $\omega_S=0$.
\section{Parameterized locked generator phases}\label{s:closedForm}
Here, we want to investigate the existence and form of stable phase-locked solutions of Eq.~\ref{e:mfG2M} for a given degree $k\geq 1$. To this end, we set $\Psi$  constant in the following and rescale time by a factor $(\lambda k)^{-1}$, so that the decoupled generator dynamics of Eq.~\ref{e:mfG2M} can be rewritten as
\begin{equation}\label{e:mfG2A}
\dot{\theta}_{\Psi,k}(z)=(\lambda k)^{-1}-f_{\Psi}(\theta,z)
\end{equation}
with $\theta_{\Psi,k}(x/k)\equiv \theta_\Psi(k,x)$, $f_\Psi(\theta,z)\equiv z\sin{(\theta-\Psi)}+(1-z)\sin{\theta}$, $\Psi\in[0,2\pi)$, integer $k\geq 1$ and $z\in[0,1]$. The contour lines $\theta^*_{\Psi,k}(z)$ defined by $f_{\Psi}[\theta^*_{\Psi,k}(z),z]=(\lambda k)^{-1}$ are called solution branches in the following, because if they encompass $z=x/k$, then $z$ yields a steady state in Eq.~\ref{e:mfG2A} and generator class $(k,x)$ is phase-locked in Eq.~\ref{e:mfG2M} for fixed $\Psi$ (stably or unstably). To understand how solution branches arise depending on $\Psi$ and $\lambda$ for given $k$, consider moreover the maxima $\hat{f}_\Psi(z)$ of curves $f_\Psi(\theta,z=\text{const})$ for each $z \in[0,1]$ given by 
\begin{equation}\label{e:fhatA}
\hat{f}_\Psi(z)\equiv\sqrt{1-2z(1-z)(1-\cos{\Psi})}\, .
\end{equation}
Clearly a solution branch encompasses $z$ as soon as 
\begin{equation}\label{e:fhatA2}
\hat{f}_\Psi(z)\geq (\lambda k)^{-1}\, .
\end{equation}
A short inspection of $\hat{f}_\Psi(z)$ confirms that if $\theta_{\Psi,k}(z\leq 1/2)$ is a steady state, so are $\theta_{\Psi,k}(1-z)$ and all $\theta_{\Psi,k}(z')$ for which $z'<z$, with the necessary coupling strength increasing the closer $z$ is to $1/2$. Hence if oscillator class $(k,x\leq k/2)$ phase-locks, so do all classes $(k,x')$ with $x'<x$ and $x'\geq k-x$, with the class of the most balanced numbers of generator and consumer neighbors phase-locking last.

Obviously $\hat{f}_{\Psi}(z)\leq 1$ always holds, so that for $\lambda<k^{-1}$, no solution branch exists for any $\Psi$. In this case, there is no steady state in Eq.~\ref{e:mfG2A} for any $z$ or $\Psi$, so that no oscillator class $(k,x)$ in Eq.~\ref{e:mfG2M} phase-locks. For larger $\lambda$, two joined solution branches $ \bar{\theta}^*_{\Psi,k}(z)$ and $\tilde{\theta}^*_{\Psi,k}(z)$ appear, related as $\tilde{\theta}^*_{\Psi,k}(z)=\pi-\bar{\theta}^*_{\Psi,k}(1-z)+\Psi$ and confined to all $z$ fulfilling $2z(1-z)(1-\cos{\Psi})\leq 1-(\lambda k)^{-2}$. Hence only generators with a sufficiently small or large number of generator neighbors $x$ phase-lock. Finally, for all coupling strengths larger than 
\begin{equation}\label{e:synchEvenA}
\lambda^*_{\Psi,k}=\left[k|\cos\left(\Psi/2\right)| \right]^{-1} \, ,
\end{equation}
both solution branches are separated, and each lives on the entire interval $z\in[0,1]$. As a consequence, all $k+1$ generator classes $(k,x)$ for a given $k$ phase-lock in Eq.~\ref{e:mfG2M}, in particular class $x=k/2$ in graphs with even-valued degree $k$. In \emph{co-evolving}  Eqs.~\ref{e:mfG2M}-\ref{e:ordG2M}, $\lambda^*_{\Psi,k}$ (computed with steady-state $\Psi$) obviously is a lower bound for the critical threshold $\lambda^*_G$ if $k$ is even-valued. 

For an odd-valued degree $k$, this lower bound is approximate: there, only the two generator classes at $x=(k-1)/2$ and $x=(k+1)/2$ must lie on the solution branches for all classes in Eq.~\ref{e:mfG2M} to phase-lock, which is the case as soon as
\begin{equation}\label{e:synchOddA}
\lambda\geq [(\lambda^{*}_{\Psi,k})^{-2}+\sin^2{\left(\Psi/2\right)}]^{-1/2}
\end{equation}
(cf. Eqs.~\ref{e:fhatA}-\ref{e:synchEvenA}). This lower bound is always smaller than $\lambda^*_{\Psi,k}$, and the better approximated by it the larger $k$ and the smaller $|\Psi|$ are. Their largest relevant difference is obtained by maximizing $\Psi$ on $[0,\pi]$ and minimizing $k$: in Eq.~\ref{e:psi2M} and Appendix~\ref{s:selfConsistent}, we observe $\Psi=\pi/2$ to be the approximately largest self-consistent stable value for various degree distributions $P(k)$ and generator abundancies $g$. Considering that $1$ is the minimum odd-valued degree, one calculates $\lambda^*_{\Psi,1}=\sqrt{2}$ versus the exact lower bound $1$ in Eq.~\ref{e:synchOddA}. Choosing the next-largest odd-valued degree $k'=3$, we already obtain $\lambda^*_{\Psi,3}=\sqrt{2}/3\approx 0.471$ versus the exact lower bound $1/\sqrt{5}\approx 0.447$.  Conversely, for the actual stable phase-locked regimes in co-evolving Eqs.~\ref{e:mfG2M}-\ref{e:ordG2M}, we observe $\lambda^*_G\geq \lambda^*_{\Psi,k'}$ in Eq.~\ref{e:psi2M} and Appendix~\ref{s:selfConsistent}, again for various $P(k)$ and $g$. Hence in the following, we assume for odd-valued degrees $k$ that $\lambda^*_G\geq\lambda^*_{\Psi,k}$ for stable steady-state $\Psi$ in co-evolving Eqs.~\ref{e:mfG2M}-\ref{e:ordG2M}. This is a weaker statement than the strict proof for even-valued $k$  that $\lambda\geq \lambda^*_{\Psi,k}$ for \emph{any} $\Psi$ in phase-locked states of Eq.~\ref{e:mfG2M}. 

Identifying $\bar{\theta}^*_{\Psi,k}(z)$ [$\tilde{\theta}^*_{\Psi,k}(z)$] with the solution branch for which $\bar{\theta}^*_{\Psi,k}(0)=0$ [$\tilde{\theta}^*_{\Psi,k}(0)=\pi$] for $\lambda\rightarrow \infty$, phase-locked states lying on $\tilde{\theta}^*_{\Psi,k}(z)$ are unstable and can thus be discarded: firstly, observe that any linearly unstable steady state in Eq.~\ref{e:mfG2M} with fixed $\Psi$ cannot be a stable phase-locked state in the system of Eqs.~\ref{e:mfG2M}-\ref{e:ordG2M} where $\Psi$ co-evolves. In turn, for oscillator classes with even-valued (odd-valued) degree $k$, Eq.~\ref{e:synchEvenA} (Eq.~\ref{e:synchOddA}) is a necessary condition for phase-locking in co-evolving Eqs.~\ref{e:mfG2M}-\ref{e:ordG2M}. Secondly, the Jacobian for Eq.~\ref{e:mfG2A} with fixed $\Psi$ and $z$ is $J_{\Psi,z}(\theta)=-\lambda \{z\cos{(\Psi-\theta)}+(1-z)\cos{\theta}\}$ and thus changes sign at the maximum of $f_{\Psi}(\theta,z=\text{const})$. As the line $\hat{f}_\Psi(z)$ of these maxima separates the two solution branches (Eq.~\ref{e:fhatA}) and furthermore $J_{\Psi,0}[\bar{\theta}^*_{\Psi,k}(0)=0]=-\lambda <0$ holds, it follows that the Jacobian is always positive for fixed points on $\tilde{\theta}^*_{\Psi,k}(z)$ [and negative for all fixed points on $\bar{\theta}^*_{\Psi,k}(z)$]. Hence the fixed points $\tilde{\theta}^*_{\Psi,k}(z)$ of Eq.~\ref{e:mfG2A} are unstable and discarded, and the stable steady states  are given by $\bar{\theta}^*_{\Psi,k}(z)$. They are of the form
\begin{eqnarray}
\sin{\left[\bar{\theta}^*_{\Psi,k}(z)\right]}&=&\frac{(\lambda k)^{-1}[1-(1-\cos{\Psi})z]+z\sin{\Psi} \sqrt{1-2z(1-z)(1-\cos{\Psi})-(\lambda k)^{-2}}}{1-2z(1-z)(1-\cos{\Psi})}\label{e:solsineA}\\
\cos{[\bar{\theta}^*_{\Psi,k}(z)]}&=&\frac{\sin{[\bar{\theta}^*_{\Psi,k}(z)]}[1-(1-\cos{\Psi})z]-(\lambda k)^{-1}}{z\sin{\Psi}}\label{e:solcosineA}\, ,
\end{eqnarray}
with $\cos{[\bar{\theta}^*_{\Psi,k}(0)]}=\sqrt{1-(\lambda k)^{-2}}$.
From the behavior of $J_{\Psi,z}(\theta)\equiv -\partial f_\Psi(\theta,z)/\partial\theta$ discussed above, it furthermore immediately follows for the contour lines $\bar{\theta}^*_{\Psi,k}(z)$ that 
\begin{equation}\label{e:slope}
\partial \bar{\theta}^*_{\Psi,k}(z)/\partial (\lambda k)\leq0 \,.
\end{equation}
\section{Self-consistent generator mean field}\label{s:selfConsistent}
A parameterized phase-locked solution $\theta^*_{-\Psi}(k,k-x)$ of Eq.~\ref{e:mfG2M} describes the same phase-locked state as $\theta^*_{\Psi}(k,x)$, but in other angular coordinates, namely for $\Psi_G\equiv 0$ and $\Psi_C\equiv-\Psi$ in Eq.~\ref{e:mfG} instead of $\Psi_C\equiv 0$ and $\Psi_G\equiv\Psi$ as before. This coordinate shift should not change the physics of generators in the system; it can however be used to simplify the computation of self-consistent $\Psi$ in Eq.~\ref{e:ordG2M}. As its left-hand side is just $\Psi_G$, Eq.~\ref{e:ordG2M} reads
\begin{equation}\label{e:psi2A}
0=\sum_{k=k_m}^{k_M}\sum_{x=0}^k w_g(k,x) \sin{[\theta^*_{-\Psi}(k,k-x)]}\equiv F_G(\Psi)
\end{equation}
in the shifted coordinates, to be fulfilled by all self-consistent stationary $\Psi$. 

A generator class with degree $k$ curtails the search space for these $\Psi$ through Eqs.~\ref{e:synchEvenA}-\ref{e:synchOddA}: with $\lambda\geq k^{-1}$ necessary for any class with degree $k$ to phase-lock (Eqs.~\ref{e:mfG2A}-\ref{e:fhatA2}), all self-consistent $\Psi$ must fulfill
\begin{equation}\label{e:intervalA}
\cos\Psi \in\begin{cases}
\left[2(\lambda  k)^{-2}-1,1\right]\\
\left[\max\{-1, 2(\lambda k)^{-2}-1-2\frac{1-(\lambda k)^{-2}}{k^2-1}\},1\right] 
\end{cases}
\end{equation}
for even- and odd-valued degree $k$, respectively. The search space of self-consistent $\Psi$ is thus dictated by the smallest realized generator degrees.

Next, we determine the linear stability of self-consistent phase-locked solutions of Eqs.~\ref{e:mfG2M}-\ref{e:ordG2M} with co-evolving mean field $\Psi$. To that end, we calculate the entries of the system's Jacobian, computed at the stationary state parametrized by $\Psi$. Consider first that with Eq.~\ref{e:ordG2M} and Sec.~\ref{s:properties}, 
\begin{equation}\label{e:psiExp}
\Psi=\text{arccot}{\left(\frac{\sum_{k=k_m}^{k_M}\sum_{x=0}^k w_g(k,x)\cos{[\theta_\Psi(k,x)]}}{\sum_{k=k_m}^{k_M}\sum_{x=0}^k w_g(k,x)\sin{[\theta_\Psi(k,x)]}}\right)}
\end{equation}
can be assumed to be well-defined sufficiently close to the fixed point. With Eqs.~\ref{e:ordG}, \ref{e:solsineM}, \ref{e:solcosineM}, \ref{e:psiExp} and some algebra, we obtain  
\begin{eqnarray*}
\frac{\partial \dot{\theta}_\Psi(k,x)}{\partial \theta_\Psi(k',x')}\Bigr|_{\substack{\theta^*}}&=&-\delta_{kk'}\delta_{xx'} \lambda\sqrt{k^2-2x(k-x)(1-\cos{\Psi})-\lambda^{-2}}\\
&&+\lambda x w_g(k',x')\cos^2{[\Psi-\theta^*_\Psi(k',x')]}(g \langle k \rangle r_G)^{-1}
\end{eqnarray*}
with
\begin{widetext}
\[\cos{[\Psi-\theta^*_\Psi(k,x)]}=\frac{[x+(k-x)\cos{\Psi}]\sqrt{k^2-2x(k-x)(1-\cos{\Psi})-\lambda^{-2}}+\lambda^{-1}(k-x)\sin{\Psi}}{k^2-2x(k-x)(1-\cos{\Psi})}\, .\]
\end{widetext}
Here $k,k'\geq 1$, $x\in[0,k]$, $x'\in[0,k']$ and $r_G$ is as in Eq.~\ref{e:ordG}. The latter reappears after having set $r_G\equiv 1$ already in assumption (a3). Methodologically, it is coherent to re-set $r_G\equiv 1$, at the cost of rendering the stability considerations inexact. The Jacobian's eigenvalues are also parameterized by $\Psi$; inserting the latter's self-consistent solutions computed in Eq.~\ref{e:psi2M} reveals the linear stability of the steady states of co-evolving Eqs.~\ref{e:mfG2M}-\ref{e:ordG2M}. If all computed eigenvalues have negative real parts, the system is linearly stable. If at least one eigenvalue possesses a positive real part, the system is linearly unstable. 
\section{Properties of locked generator phases}\label{s:properties}
First, we show for fixed even-valued degree $k$ that if $\Psi\in[0,\pi)$, then $\theta^*_\Psi(k,x)\in[0,\pi)$ for all $x\in[0,k]$ in Eq.~\ref{e:mfG2M}. According to Eq.~\ref{e:solsineM}, this is obviously the case for $\Psi=0$. If $\Psi\in(0,\pi)$, consider that for each  $x$,
\[k^2-2x(k-x)(1-\cos{\Psi})-\lambda^{-2}\geq 0\]
holds through Eqs.~\ref{e:fhatA}-\ref{e:fhatA2}. According to Eq.~\ref{e:solsineM}, it is then sufficient to show that 
\begin{eqnarray}\label{e:ineqA}
0&<&\lambda^{-1}[k-(1-\cos{\Psi})x]+ x\sin{\Psi} \sqrt{k^2-2x(k-x)(1-\cos{\Psi})-\lambda^{-2}}
\end{eqnarray}
for $\Psi\in(0,\pi)$ and all $x\in[0,k]$. Equation \ref{e:ineqA} is surely fulfilled for all integer $x< k(1-\cos{\Psi})^{-1}$. For all larger $x$, Eq.~\ref{e:ineqA} reduces to $x>(\lambda \sin{\Psi})^{-1}$. By proposition, this is surely fulfilled if $k(1-\cos{\Psi})^{-1}> (\lambda \sin{\Psi})^{-1}$. As $\lambda\geq\lambda^*_{\Psi,k}$ in all phase-locked regimes of Eq.~\ref{e:mfG2M} for all even-valued $k$ (Eq.~\ref{e:synchEvenA}), this is in turn surely true if $\lambda^*_{\Psi,k}> (1-\cos{\Psi})(k\sin{\Psi})^{-1}$. Again with Eq.~\ref{e:synchEvenA} for even-valued $k$, this last inequality reduces to $1> \cos{\Psi}$, which is true for all $\Psi\in(0,\pi)$. Hence $\sin{[\theta^*_\Psi(k,x)]}> 0$ in Eq.~\ref{e:solsineM} and thus
$\theta^*_\Psi(k,x)\in[0,\pi)$ for all $\in[0,k]$ and $\Psi\in[0,\pi)$  if $k$ is even-valued.

This also applies to all generator classes with degree $k'>k$, where $k'$ can be even- or odd-valued. The reason is a smaller upper bound for $\theta^*_\Psi(k',x)$ through Eq.~\ref{e:slope} and a positive lower bound through $\sin{[\theta^*_\Psi(k,x)]}=z\sin{\Psi}/\hat{f}(z)\geq 0$ for $\lambda\rightarrow \infty$ (Eqs.~\ref{e:fhatA}, \ref{e:solsineA} and \ref{e:slope}). From $\theta^*_\Psi(k,x)\in[0,\pi)$ for $\Psi\in[0,\pi)$ follows $\sin{[\theta^*_{-\Psi,k}(k,k-x)]}> 0$ for $\Psi\in(-\pi,0]$. This inequality applies to all generator classes $k'>k$, so that if the smallest realized generator degree $k_m$ in the network is even-valued, Eq.~\ref{e:psi2M} cannot be fulfilled for any $\Psi\in(-\pi,0]$. Hence in that case, any self-consistent $\Psi$ must lie in the interval $(0,\pi)$ (as $\Psi=\pi$ can be immediately ruled out with Eqs.~\ref{e:solsineM} and \ref{e:psi2M}).

If however the smallest realized generator degree $k_m$ in the network is odd-valued, then crucially $\lambda\geq \lambda^*_{\Psi,k_m}$ is not true for all $\Psi\in[0,\pi)$ in supercritical regimes of Eq.~\ref{e:mfG2A}. More specifically, it does not hold for $\Psi$ for which $|\Psi|\sim \pi/2$. Hence the contribution $\sum_{x=0}^{k_m} w_g(k_m,x)\sin{[\theta^*_{-\Psi,k}(k_m,k_m-x)]}$ to the sum in Eq.~\ref{e:psi2M} can be negative for those $\Psi$. If on the one hand this negative contribution is small, as to be expected in heterogeneous network topologies where $P(k_m)\ll 1$, then above considerations still hold. If on the other hand $P(k_m)\approx 1$ , as for example in regular random graphs, then above considerations still hold for all $\lambda_G\geq \lambda$, as $\lambda^*_G\geq \lambda^*_{\Psi,k_m}$ is assumed in phase-locked regimes of co-evolving Eqs.~\ref{e:mfG2M}-\ref{e:ordG2M} (cf. Appendix~\ref{s:closedForm}).

Our strict reasoning in the case of even-valued smallest generator degrees $k_m$ and our reasonable assumption in the case of an odd-valued $k_m$ leads to the following conclusions about linearly stable phase-locked states in the co-evolving Eqs.~\ref{e:mfG2M}-\ref{e:ordG2M}: 

\textbf{(c1)} $\Psi$ lies in the intersection of $(0,\pi)$ with the smallest interval defined in Eq.~\ref{e:intervalA}.

\textbf{(c2)} As $\sin{[\theta^*_{\Psi}(k,x)]}>0$ on $\Psi\in(0,\pi)$, it follows from (c1) that $\theta^*_{\Psi}(k,x)\in(0,\pi)$. In particular, the locked phases of generator classes are all less than $\pi$ apart from each other. 

\textbf{(c3)} Locked phases can be written as $\theta^*_{\Psi}(k,x)=\textrm{arccot}\left(\cos{[\theta^*_{\Psi}(k,x)]}/\sin{[\theta^*_{\Psi}(k,x)]}\right)$ due to (c2), so that with Eqs.~\ref{e:solsineA}-\ref{e:solcosineA}, $d\bar{\theta}^*_{\Psi,k}(z)/dz\geq 0$ for all $z\in[0,1]$ and thus $\theta^*_\Psi(k,x+1)\geq\theta^*_\Psi(k,x)$ for all $k\geq 1$ and $x\in[0,k-1]$. 

\textbf{(c4)}
As a consequence of (c2)-(c3), the maximum difference on $(0,\pi)$ between any two generator phases is, for fixed $k$ and $\Psi$, $\Delta_{\Psi,k}\equiv\theta^*_{\Psi}(k,k)-\theta^*_{\Psi}(k,0)$.  Obviously $\theta^*_{\Psi}(k,0)=\arcsin{(\lambda k)^{-1}}$ (Eqs.~\ref{e:solsineM}-\ref{e:solcosineM}), while $\sin{[\theta^*_\Psi(k,k)-\Psi]}=\arcsin{(\lambda k)^{-1}}$ (Eq.~\ref{e:solcosineM}) leads to  $\theta^*_\Psi(k,k)=\arcsin{(\lambda k)^{-1}}+\Psi$ (Eq.~\ref{e:solsineM}) and finally to $\Delta_{\Psi,k}=\Psi$, an upper bound for the maximum phase difference between any two locked generator phases. Moreover, it immediately follows that $\theta^*_\Psi(k,x)\in[\arcsin{(\lambda k)^{-1}},\Psi+\arcsin{(\lambda k)^{-1}}]$.

\textbf{(c5)}
From Eq.~\ref{e:slope} follows directly that the larger a generator's degree $k$ with constant neighborhood composition $x/k$ is, the smaller is its locked phase on $(0,\pi)$.

\section{Links to other analytic approaches}\label{s:generalization}
In regular random graphs with the same degree $k$ \emph{and} neighborhood composition for all oscillators, $(k,kg)$ with an integer $k  g$ is the only relevant oscillator class. Hence in these highly regular settings, $w_g(k',x')=k  g  [\delta(k-k')\delta(k  g-x')]$ with the Dirac $\delta$ function, so that  Eqs.~\ref{e:ordG}-\ref{e:ordC} imply $\theta_Y(k,k   g)=\Psi_Y=\hat{\Psi}_Y$ at all times for oscillator type $Y\in\{G,C\}$. Therefore the system reduces to an effective two-oscillator problem. For both decoupled generator and consumer dynamics, Eqs.~\ref{e:solsineM} and \ref{e:psi2M} (and their consumer counterparts) yield $\sin \Psi=[\lambda k (1-g)]^{-1}$ for the respective mean-field parameter. In both cases, this parameter is trivially self-consistent as then $F_{G,C}(\Psi)\equiv 0$. Hence $\Psi^*_G=\hat{\psi}^*_C$ and also $\hat{\Psi}^*_G=\psi^*_C$ for locked neighborhood phases, so that locked consumer phases can be unambiguously expressed by generator coordinates. In these coordinates, the difference $\Psi=\Psi_G-\Psi_C$ between locked generator and consumer phases is then the relevant system variable. Its linear stability analysis in Appendix~\ref{s:selfConsistent} then reduces to determining when $1-2 g(1- g)(1-\cos{\Psi})\geq (\lambda k)^{-2}$. Considering that $g\in(0,1/2]$ and $k>0$, this reveals a unique stable state at $\Psi=\arcsin\left([\lambda k (1-g)]^{-1}\right)$ above the critical threshold $\lambda^*= [ k(1-g)]^{-1}$.

The same results are obtained by a more naive mean-field approximation of Eq.~\ref{e:kuramoto}. There, oscillators $j$ of same natural frequency can be considered equivalent through setting $\theta_j=\Psi_G$ for $j\in[1,g  N]$ and $\theta_j=\Psi_C$ for $j\in[g N+1,N]$. Splitting the coupling term into two contributions from interactions with $k  g$ generators and with $k (1-g)$ consumers, as well as averaging Eq.~\ref{e:kuramoto} over the generator and consumer ensemble yields $\dot{\Psi}=(1-g)^{-1}-\lambda k \sin{\Psi}$ for the phase difference $\Psi\equiv\Psi_G-\Psi_C$, with steady states and stability as above.
\newpage
\renewcommand\refname{\textbf{References}}
\bibliographystyle{unsrt}
\bibliography{draft}
\end{document}